\documentclass[preprint,12pt,3p,authoryear]{elsarticle}
\setcitestyle{square} 

\usepackage{amssymb}
\usepackage{amsmath}
\usepackage{xcolor}
\usepackage{color}
\usepackage{float}

\journal{Nuclear Instruments and Methods in Physics Research, B}

\setcitestyle{numbers}

\begin{document}

\begin{frontmatter}


\title{Dark Matter Detector Radioimpurities $^{129}$I and $^{210}$Pb Measured with Accelerator Mass Spectrometry} 

\author[ANU,ARC]{Zuzana Slavkovsk\'{a} \corref{email}}
\author[ANU,ARC,UCL]{Ferdos Dastgiri} 
\author[ANU]{L.Keith\,Fifield} 
\author[ANU,ARC]{Michaela\,B.\,Froehlich} 
\author[ANSTO]{Michael\,A.C.\,Hotchkis} 
\author[ANU,ARC,HZDR]{Dominik Koll} 
\author[HZDR,UV]{Silke Merchel} 
\author[ANU]{Stefan Pavetich} 
\author[ANU,ARC]{Stephen\,G.\,Tims} 
\author[ANU,ARC,HZDR]{Anton Wallner} 

\cortext[email]{Corresponding Author\\ 
\indent ~~\textit{Email address: zuzana.slavkovska@anu.edu.au} (Z.\,Slavkovsk\'{a}).}

\affiliation[ANU]{organization={Department of Nuclear Physics and Accelerator Applications, Research School of Physics, The Australian National University}, 
            city={Canberra},
            postcode={2601}, 
            state={ACT},
            country={Australia}}
            
\affiliation[ARC]{organization={ARC Centre of Excellence for Dark Matter Particle Physics},
            country={Australia}}

\affiliation[ANSTO]{organization={Centre for Accelerator Science, Australian Nuclear Science and Technology Organisation},
           state={NSW},
           city={Sydney},
            postcode={2234},
            country={Australia}}

\affiliation[HZDR]{organization={Helmholtz-Zentrum Dresden-Rossendorf, Institute of Ion Beam Physics and Materials Research},
            city={Dresden},
            postcode={01328},
            country={Germany}}

\fntext[UCL]{Now:\,Department of Physics and Astronomy, University College London, London, WC1E 6BT, UK}

\fntext[UV]{Now:\,University of Vienna, Faculty of Physics, Isotope Physics, Vienna, 1090 Austria}

\begin{abstract}
Sodium iodide crystals doped with thallium NaI(Tl) can be used as detector material for direct dark matter detection by taking advantage of their particle detection properties of scintillation. In order to achieve this, it is crucial that these crystals are of ultra-high purity. Radioimpurities within the crystals may potentially mimic dark matter signals and thus must be quantified, minimised where possible and distinguished from real events. Abundances of radionuclides $^{129}$I and $^{210}$Pb, which are dominant sources of radioimpurities in NaI(Tl) crystals, were measured using accelerator mass spectrometry at the Australian National University (ANU) and the Australian Nuclear Science and Technology Organisation (ANSTO). NaI powder chemically processed to AgI, and, for the first time, unprocessed NaI powder, were shown to be suitable as AMS targets. A consistent $^{129}$I/$^{127}$I ratio of (2.0\,$\pm$\,0.3)\,x\,10$^{-13}$ was measured in three different commercially available NaI powders. Therefore, it was concluded that the choice of NaI powder has a negligible influence on the $^{129}$I contribution to low-background dark matter experiments. For $^{210}$Pb, different Pb molecular ion species were assessed with PbO$_2^-$ being the preferred species and applied to investigate different lead oxide compounds for their suitability as Pb carriers. A $^{210}$Pb/Pb isotopic ratio of (3.6 $\pm$ $^{1.7}_{1.4}$)\,x\,10$^{-15}$ was measured in Pb$_3$O$_4$ powder. This met the required lower activity limit of $^{210}$Pb when adding 1\,mg of stable lead into 1\,kg of NaI(Tl) powder with a desired maximum $^{210}$Pb/Pb isotopic ratio of 1\,x\,10$^{-14}$. These results indicate the suitability of the investigated Pb$_3$O$_4$ as a potential carrier for incorporation with Pb extracted from NaI used for dark matter experiments.

\end{abstract}


\begin{keyword}

Accelerator mass spectrometry 
\sep Dark matter
\sep $^{129}$I 
\sep $^{210}$Pb

\end{keyword}

\end{frontmatter}


\section{Introduction}
\label{sec1}
Observations of gravitational interactions between astronomical entities indicate the presence of dark matter throughout the universe \cite{Bertone}. Due to the orbit of the Earth around the Sun, an annual modulation of dark matter flux on Earth is expected \cite{modulation}. Such modulation has been observed for over 20 years by the DAMA/LIBRA (DArk MAtter/Large sodium Iodide Bulk for RAre processes) experiment \cite{dama_libra}. A potential dark matter particle candidate is the weakly interacting massive particle (WIMP) \cite{wimps}. DAMA/LIBRA is located at the Laboratori Nazionali del Gran Sasso (LNGS) in Italy and consists of an array of high-purity NaI(Tl) crystals with a total mass of about 250\,kg. The observed signal is in the energy region 2-6\,keV and has a statistical significance of 12.9\,$\sigma$; however, the modulation signal observed by DAMA/LIBRA has neither been reproduced nor ruled out by any other model-independent experiment worldwide. In particular, no other NaI(Tl)-based dark matter experiments (such as COSINE-100 \cite{cosine} and ANAIS \cite{anais}) have been able to reproduce this modulation. To investigate this, construction of SABRE (Sodium iodide with Active Background REjection \cite{background}) South, the first NaI(Tl) crystal underground dark matter direct detection experiment in the southern hemisphere, has been initiated in Australia by the Australian Research Council (ARC) Centre of Excellence for Dark Matter Particle Physics. Locating SABRE in the first underground southern hemisphere laboratory, the Stawell Underground Physics Laboratory (SUPL) in Australia will permit any seasonal effects impacting the modulation signal observed by DAMA/LIBRA to be conclusively measured. In addition, the SABRE program has also planned the construction of a NaI(Tl) detector in the northern hemisphere located at LNGS in Italy \cite{background}. This twin experiment provides an unprecedented opportunity to investigate the modulation signal, indicating a possible dark matter signal if it is in phase, and a seasonal effect if it is out of phase.  
\\
\\
Radiation emitted by naturally occurring radionuclides in dark matter detector materials with half-lives in the order of years to billions of years induce a background activity underlying all measurements. It is therefore critical to minimise and quantify any radioactive impurities within the NaI(Tl) crystals to account for interferences with a positive dark matter signal. 
\\
\\
Furthermore, signal contamination caused by production of new radionuclides in the crystal by cosmic radiation may also interfere with dark matter signals. This can be minimised by placing the detector underground. Problematic radionuclides produced by cosmic radiation \cite{background} are often shorter-lived with half-lives in the order of days. A 'cool-down' period of six months from the placement of the crystals underground is generally carried out to minimise this contamination. A time-dependent signature of this decaying background signal can be measured in order to determine the overall contribution of the cosmogenic detector background. 
\\
\\
Of particular interest are the radionuclides $^{129}$I and $^{210}$Pb as they are dominant background sources in the NaI(Tl) crystals used for direct dark matter detection \cite{background}. With half-lives in the order of years to megayears, the activities of these radionuclides cannot be removed and must instead be quantified to calibrate the measurements. A small amount of $^{129}$I is produced naturally from spontaneous fission of natural uranium and from spallation of xenon in the Earth's atmosphere by cosmic ray muons; however, the majority of $^{129}$I arises from anthropogenic contamination from nuclear weapons testing, nuclear reactor incidents and nuclear fuel reprocessing. Iodine-129 with a half-life of 16.14 (12)\,Myr \cite{garcia_torano}, decays to $^{129}$Xe via beta decay, to an excited level at 39.6\,keV and to the ground state with probabilities of 99.5\% and 0.5\%, respectively \cite{garcia_torano}. The activity of $^{129}$I can be determined by neutron activation with subsequent decay counting, or deduced from measurements of the $^{129}$I content, using accelerator mass spectrometry (AMS) as presented in this study. 
\\
\\
Lead-210 is a part of the $^{238}$U decay chain and therefore occurs naturally in the environment. It has a half-life of 22.2 years and the radiation associated with its beta-decay to the ground state of $^{210}$Bi overlaps with the low-energy region, in which the dark matter signal is expected. Lead-210 is routinely measured by decay counting \cite{kristensen} and inductively coupled plasma mass spectrometry; however, advances in AMS techniques allow the measurement of $^{210}$Pb at the required level through direct atom counting \cite{scavenger, vivo-vilches}. These advances have been achieved using the 14\,UD tandem accelerator \cite{wallner} at the ANU in Canberra and the 1\,MV VEGA accelerator \cite{hotchkis} at ANSTO in Sydney. 

\section{Radioimpurity $^{129}$I}

Iodine-129 is routinely measured using AMS. Isobaric interference is negligible as Xe does not form negative ions. Typically, the isotopic ratio $^{129}$I/$^{127}$I is measured \cite{roberts}. Iodine-129 is incorporated within the NaI matrix and is therefore a natural part of the NaI(Tl) powder used to grow the detector crystals for direct dark matter detection. For this work, three types of commercial NaI powder were analysed by AMS: (i)\,astro-grade quality powder from Sigma-Aldrich (Merck), (ii)\,growth-grade quality powder from Sigma-Aldrich (Merck) and (iii)\,an ANU-in-house analytical-grade powder from May and Baker Ltd. The former is the purest commercially available NaI powder with a purity of 99.99\,(5)\%.

\subsection{Sample Preparation}
Three sample materials from the three NaI powders were trialed to establish optimised conditions for $^{129}$I AMS measurements: (i)\,commercial NaI powder directly as purchased, with no chemical processing, (ii)\,commercial NaI powder mixed with Ag powder (Goodfellow, 99.99\%, 100\,$\mu$m) and a (iii)\,NaI chemically converted to AgI powder mixed with Ag. The mixing with Ag was at a 1:1 ratio by weight. To produce AgI, sodium iodide was combined with solution of AgNO$_3$ to precipitate silver iodide. The produced sample was centrifuged, the sodium nitrate decanted and the AgI dried at 90$^{\circ}$C. 
\\
\\
Additionally, blank material known as Woodward iodine in the form of AgI with a known $^{129}$I/$^{127}$I isotopic ratio of 1.3\,x\,10$^{-14}$ \cite{keith} and reference material referred to as 'Kuni Standard' with a ratio of 2.68\,x\,10$^{-12}$ were measured and used to normalise the results of the investigated powders. Both were provided by Prof.\,June Fabryka-Martin and prepared at the University of Arizona in the early 1990s.

\subsection{AMS Measurements}
The AMS measurements of $^{129}$I were performed at the Heavy Ion Accelerator Facility (HIAF) at the ANU using the 14\,UD pelletron tandem accelerator \cite{dracoulis}. The samples were mounted into the multi-cathode source of negative ions by caesium sputtering (MC-SNICS) and a terminal voltage of 11\,MV was applied to the accelerator. The sample was sputtered with Cs$^+$ ions. Negative ions from the sample were pre-accelerated to 155\,keV, mass analysed with a 90$^{\circ}$ magnet, and atomic masses 127 or 129 were injected into the accelerator. In the terminal, molecular ions were dissociated with a nitrogen gas stripper, and $^{127}$I or $^{129}$I with a charge state of 7+ was then selected by the analysing magnet. The transmission of the 7+ charge state was 6.5\%. 
\\
\\
Two methods of detection were applied: (i)\,time-of-flight (TOF) with a 6\,m-long flight path together with a Si-detector with an active area of 25\,mm in diameter \cite{Fifield2008, Wilcken} and (ii)\,Si-detector only. There was no evidence of a $^{127}$I background, so the TOF information was not required and the Si detector was used for detector-only measurements at higher detector efficiency.  
\\
\\
The NaI and AgI sample materials delivered stable $^{127}$I beams of approximately 5\,$\mu$A, which were reliably attenuated by a factor of 100 with an electrostatic chopper in order to eliminate the effect of loading on the terminal voltage, which provided better terminal voltage stability. The injected $^{129}$I beams were not attenuated. Results indicated no enhanced background when adding Ag powder in comparison to the pure material. As mixing with Ag increases electrical and thermal conductivity of the sample and results in a more stable ion source output compared to samples without Ag powder, measurements evaluated in this work were performed using NaI and AgI, both mixed with Ag. 
\\
\\ 
Table \ref{tab1} shows the $^{129}$I/$^{127}$I ratio of the three materials investigated here. Uncertainties are quoted as the larger of the internal and external errors. The ratios are in agreement with each other, and with previous measurements at the ANU \cite{keith}. Furthermore, our results are in reasonable agreement with the DAMA/LIBRA ratio \cite{bernabei2008}. 
\\
\\
The consistency of the $^{129}$I/$^{127}$I ratios implies that the commercially available NaI investigated here has a similar origin. For this reason, there is no distinction in quality between different NaI powders for use within low background dark matter experiments, as they all contain an equal amount of $^{129}$I.
\\
\\

\begin{table}[H]
\caption{$^{129}$I/$^{127}$I ratios measured by AMS at the HIAF at the ANU, the corresponding specific $^{129}$I activities per kg NaI and number of $^{129}$I atoms per kg of different types of NaI, compared to the DAMA/LIBRA value \cite{bernabei2008}.} 
\vspace{0.3 cm}
\label{tab1}
\centering
\begin{tabular}{l| c| c| c}
NaI powder type& Ratio $^{129}$I/$^{127}$I & Activity  &  $^{129}$I atoms/kg          \\
   & & [mBq/kg] & (10$^{11}$\,at) \\
  \hline
Astro-grade    & (1.93 $\pm$  0.22)\,x\,10$^{-13}$   & 1.06 $\pm$ 0.12 & 7.8 $\pm$ 0.9  \\
Growth-grade   & (2.2 $\pm$   0.4)\,x\,10$^{-13}$    & 1.23 $\pm$ 0.22 & 9.0 $\pm$ 1.6  \\
ANU in-house   &  (1.98 $\pm$  0.21)\,x\,10$^{-13}$  & 1.09 $\pm$ 0.12 & 8.0 $\pm$ 0.9  \\
DAMA/LIBRA\cite{bernabei2008}  &(1.7 $\pm$ 0.1)\,x\,10$^{-13}$     & 0.93 $\pm$ 0.05    & 6.8 $\pm$ 0.4\\
\end{tabular}

\end{table}

\section{Radioimpurity $^{210}$Pb}
Ultra-pure NaI(Tl) powder is commercially available at high cost and in low quantities limiting our radiopurity analysis to not more than a kilogram of NaI. The stable lead concentration in this amount of powder is not sufficient to produce the required AMS sample of at least a milligram. Therefore, it is necessary to add a lead carrier with negligible $^{210}$Pb content to produce a sufficient amount of lead.  
\\
\\
The expected $^{210}$Pb concentration in dark matter NaI(Tl) crystals is 0.03\,mBq/kg NaI \cite{antonello}, which corresponds to a $^{210}$Pb/$^{208}$Pb isotopic ratio of 2\,x\,10$^{-14}$ (or a $^{210}$Pb/Pb ratio of 1\,x\,10$^{-14}$) when 1\,mg of stable lead carrier is added to 1\,kg of NaI(Tl). The ratio of the Pb carrier must be well-known and lower than this value.
\\
\\
Potential carrier materials were analysed at the ANU and at ANSTO using AMS in order to find a lead carrier material with a suitably low $^{210}$Pb content.

\subsection{Sample Preparation}
Three different lead compounds that were available at the ANU were investigated as AMS target materials: (i) Pb$_3$O$_4$ (The British Drug Houses Ltd.), (ii) PbO$_2$ (The British Drug Houses Ltd.) and (iii) PbO (origin unknown). The three different oxides above were measured either as pure oxides or were mixed with Ag powder (Goodfellow, 99.99\%, 100\,$\mu$m) at a 1:1 mixing ratio by weight to improve electrical and thermal sample conductivity. Negative ion output studies of lead oxides and fluorides have been previously performed at the ANU \cite{scavenger}.
\\
\\
In addition, reference samples with a $^{210}$Pb/Pb isotopic ratio of 1.0\,x\,10$^{-10}$ and 1\,x\,10$^{-11}$ were produced using a $^{210}$Pb solution sourced from the Environmental Research Institute of the Supervising Scientist (ERISS), provided by Peter Medley. This solution with a known activity of 142.8\,$\pm$\,3.2 Bq/g (as of March 25th, 2014) was combined with a Pb solution produced from stable metallic lead powder (Alfa Aesar, 99.9\%, 200 mesh) and processed to lead oxide by heating it to 200$^\circ$C for 8 hours to produce the reference materials (see Table \ref{lead_std}). The measured $^{210}$Pb/$^{208}$Pb isotopic ratios of these samples were used to normalise the results of the different chemical lead compounds analysed in this work.

\begin{table}[H]
\caption{Material composition of the reference lead samples measured at the VEGA facility at ANSTO and their nominal $^{210}$Pb/Pb ratio. The samples were produced using the ERISS $^{210}$Pb solution.}
\label{lead_std}
\vspace{0.3 cm}
\centering
\begin{tabular}{ c| c | c}
 Material origin & Sample composition  & $^{210}$Pb/Pb ratio        \\
    \hline
Reference A1       & 2.4\,mg PbO + 4.7\,mg Ag  & 1\,x\,10$^{-10}$  \\
Reference A2       & 2.0\,mg PbO + 3.8\,mg Ag  & 1\,x\,10$^{-10}$  \\
Reference A3       & 4.2\,mg PbO               & 1\,x\,10$^{-10}$  \\
Reference B\,\,\,  & 1.7\,mg PbO + 3.5\,mg Ag  & 1\,x\,10$^{-11}$  \\
\end{tabular}
\end{table}

\subsection{AMS Measurements}
\begin{figure}[H]
\centering
\includegraphics[width=\textwidth]{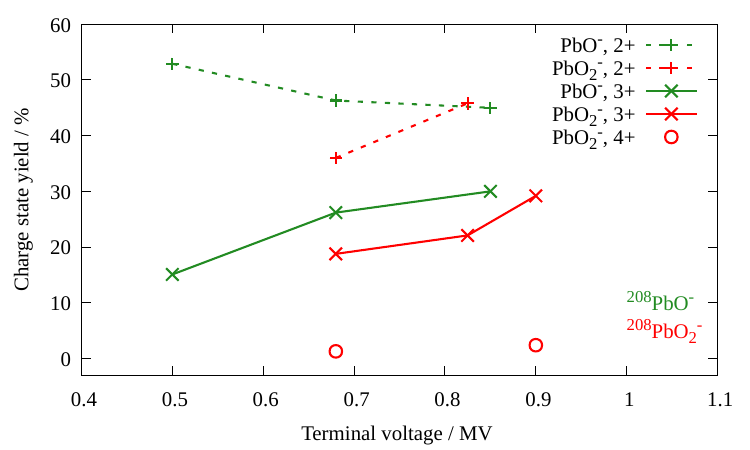}
\caption{Comparison of charge state yields at terminal voltages between 0.5\,MV and 0.9\,MV for charge states 2+, 3+ and 4+ for extracted $^{208}$PbO$^-$ and $^{208}$PbO$_2^-$ ions. The measurements were performed at the VEGA accelerator at ANSTO.}
\label{chargestate}
\end{figure}

Lead-210 is not routinely measured with AMS. Its isobars are unstable $^{210}$Bi and $^{210}$Po; however, both are expected to have significantly lower concentrations in seculiar equilibrium than $^{210}$Pb. A study of the measurement background as well as the measurement efficiency was performed to ensure that $^{210}$Pb measurements with AMS are feasible in addition to finding a carrier material with a sufficiently low $^{210}$Pb concentration. 
\\
\\
The AMS measurements of $^{210}$Pb were conducted with the 1\,MV VEGA accelerator at ANSTO \cite{hotchkis}. Additional beam output tests were conducted at HIAF at the ANU with the same ion source and injection magnet as used for AMS with the 14\,UD accelerator. Both facilities utilise an MC-SNICS with 134 and 32 sample positions, respectively. 
\\
\\
In order to determine the optimal accelerating voltage for the highest measurement efficiency, the charge state yields when injecting molecular lead anion species PbO$^-$ and PbO$_2^-$ were investigated at the VEGA facility at a range of terminal voltages between 0.5\,MV and 0.9\,MV using helium stripping gas. Charge states 2+, 3+ and 4+ were tested. The results are listed  in Table \ref{charge_state_numbers} and shown in Figure \ref{chargestate}. The analysing magnet bending power limit was exceeded for the 2+ charge state at 0.9\,MV. Overall, charge state 2+ shows the highest charge state yield, with 3+ ranging from two to three times lower and 4+ being low at just a few percent. PbO${^-}$ ions show a slightly higher yield than PbO${_2^-}$ ions by a few percent, with the difference decreasing towards higher terminal voltages. The obtained results were compared with the results from \cite{ottawa}, in which the ion yields at a terminal voltage of 1.45\,MV have been investigated at the IsoTrace AMS Laboratory at the University of Toronto \cite{ottawa}. These data confirm the trend observed in this work of lower charge states having the highest yield and decreasing towards higher charge states.

    \begin{table}[H]
    \caption{Comparison of the charge state yields for $^{208}$Pb at terminal voltages between 0.5\,MV and 0.9\,MV for charge states 2+, 3+ and 4+ for extracted $^{208}$PbO$^-$ and $^{208}$PbO$_2^-$ ions. The measurements were performed at the 1\,MV VEGA facility at ANSTO using helium stripping gas.}
\label{charge_state_numbers}
\vspace{0.3 cm}
\centering
\begin{tabular}{c| c c c| c c c}
Terminal & \multicolumn{3}{c}{$^{208}$PbO$^-$} & \multicolumn{3}{c}{$^{208}$PbO$_2^-$} \\
 voltage [MV]& 2+    &  3+  &  4+  &  2+    &  3+    &  4+    \\
\hline
0.50  & 52.9\%      & 15.1\%    & -         &  -        &   -       &   -               \\
0.68  & 46.3\%      & 26.2\%    &  -        &  36.0\%   & 18.8\%    &    1.3\%          \\
0.83  & -           &  -        & -         &  45.8\%   & 22.1\%    &  -                 \\
0.85  &  45.0\%     &  30.0\%   & -         & -         &    -      &                   \\ 
0.90  & -           &  -        & -         & -         &  29.2\%   & 2.4\%             \\
\end{tabular}
\end{table}

\begin{figure}[H]
\centering
\includegraphics[width=\textwidth]{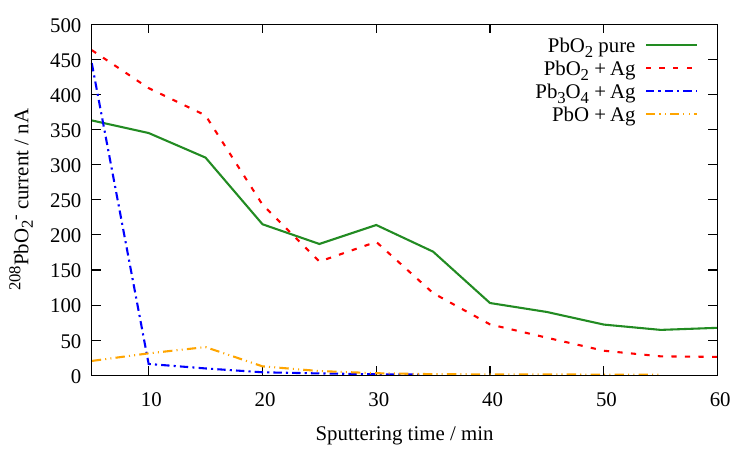}
\caption{Time evolution of $^{208}$PbO$_2^-$ currents from Pb$_3$O$_4$, PbO$_2$ and PbO mixed with Ag powder with a mixing ratio of 1:1 by weight, and PbO$_2$ pure. The samples were measured at ANSTO using helium stripping gas.}
\label{currents}
\end{figure}

Although the 2+ state has a higher yield, the 3+ charge state was chosen for the measurements reported here to avoid possible interferences from 2+ molecular ions that may have survived the stripping process. In oder to maximise the 3+ yield from the helium stripper, a terminal voltage of 0.9\,MV was used. Furthermore, the PbO$_2^-$ current from the ion source is usually higher than PbO$^-$, making the overall efficiency of PbO$_2^-$ favourable to PbO$^-$ and making PbO$_2^-$ the preferred molecular ion. With PbO$_2^-$ ions extracted from the ion source, the charge state yield of $^{208}$Pb$^{3+}$ ions was 29.2\%. The measurement cycle consisted of 300\,ms for counting $^{210}$Pb in the gas-ionisation detector and 3\,ms each for the $^{206}$Pb and $^{208}$Pb currents. Each run consisted of 1000 cycles. 
\\
\\
Figure \ref{currents} shows the course of the stable $^{208}$Pb current during the measurements for Pb$_3$O$_4$, PbO$_2$ and PbO powders. The PbO$_2$ samples mixed with Ag powder delivered the highest $^{208}$PbO$_2$ beam current of about 470\,nA at the beginning of the measurement, followed by Pb$_3$O$_4$ with about 440\,nA and pure PbO$_2$ with 370\,nA. The current of PbO mixed with Ag was one order of magnitude lower, starting at 20\,nA. Pb$_3$O$_4$ was sputtered relatively quickly, on average within 10 minutes, while the current of the PbO$_2$ material, both pure and mixed with Ag, dropped by a factor of 10 over a sputtering time of one hour. The currents from the PbO powder material decreased to $<$10\,nA within 20 minutes.

Compared to Pb$_3$O$_4$ and PbO$_2$, reference material samples delivered relatively weak stable $^{208}$Pb beam currents starting between 20\,nA and 100\,nA for $^{208}$PbO$_2^-$. This correlates with the trend in Figure \ref{currents}, as the reference samples are in the form of PbO powder. 
\\
\\
The negative ion formation for $^{208}$Pb for extracted PbO$_2^-$ ions on the low-energy side of VEGA at ANSTO was 4.4\,x\,10$^{-3}$, and at HIAF at the ANU 3\,x\,10$^{-3}$. The overall measurement efficiency of VEGA for $^{208}$Pb for the charge state 3+ was 1.3\,x\,10$^{-3}$.
\\
\\
The determined $^{210}$Pb/Pb isotopic ratios, the corresponding specific $^{210}$Pb activities in 1\,mg Pb and the number of $^{210}$Pb atoms in 1\,mg Pb are listed in Table \ref{ratios_lead}. The larger of the internal and external uncertainties are quoted as uncertainties. Uncertainties for low counting statistics $<$20 were calculated using asymmetric confidence intervals according to Feldman and Cousins \cite{feldman_cousins} with 1-$\sigma$ confidence levels.  The investigated Pb$_3$O$_4$ powder has a ratio that, within the uncertainties, satisfies the condition of having a lower $^{210}$Pb concentration than expected in dark matter NaI(Tl) crystals, corresponding to 0.03\,mBq/kg \cite{antonello} with a $^{210}$Pb/Pb isotopic ratio of 1\,x\,10$^{-14}$.  PbO$_2$ and PbO do not satisfy the maximum activity condition within their uncertainties.

\begin{table}[H]
\caption{$^{210}$Pb/Pb ratios measured by AMS at the 1\,MV VEGA accelerator at ANSTO, the corresponding specific $^{210}$Pb activities in 1\,mg Pb, and the number of $^{210}$Pb atoms in 1\,mg Pb.}
\label{ratios_lead}
\vspace{0.3 cm}
\begin{tabular}{c |l| c |c}
Type of & Ratio $^{210}$Pb/Pb & Activity & $^{210}$Pb atoms/g\,Pb\\   
Pb powder & & [mBq/g\,Pb] & (10$^7$\,at)\\
   \hline
\rule{0pt}{3ex}   
Pb$_3$O$_4$     &(3.6 $\pm$ $^{1.7}_{1.4}$)\,x\,10$^{-15}$          & 10.3 $\pm$ $^{4.9}_{4.0}$         & 1.0 $\pm$ $^{0.5}_{0.4}$  \\
\rule{0pt}{3ex}   
PbO$_2$         &(7.9 $\pm$  2.2)\,x\,10$^{-15}$                    & 22.5 $\pm$ 6.4                    & 2.3 $\pm$ 6.5             \\
\rule{0pt}{3ex}   
PbO             &(8.2 $\pm$ $^{14.4}_{\,\,\,5.2}$)\,x\,10$^{-15}$   & 23.5 $\pm$ $^{41.1}_{14.8}$       & 2.4 $\pm$ $^{4.2}_{1.5}$  \\
\end{tabular}
\end{table}

\section{Summary and Conclusions}
Quantifying iodine and lead radioisotope impurities by measuring $^{129}$I and $^{210}$Pb in dark matter detector materials is crucial for the success of dark matter experiments. Here, both radioisotope abundances we have measured wth AMS at the 14\,UD HIAF at the ANU and at the 1\,MV VEGA accelerator at ANSTO.
\\
\\
To quantify $^{129}$I levels, $^{129}$I/$^{127}$I isotopic ratios of astro-grade quality, growth-grade quality and an ANU in-house analytical-grade NaI powder were measured. NaI powder as purchased as well as chemically processed to AgI, both as pure powders and mixed with Ag, delivered stable currents of about 5\,$\mu$A. The transmission of the investigated 7+ state measured at the HIAF at the ANU was 6.5\%. Two measurement modes were used: time-of-flight and detector-only. A consistent $^{129}$I/$^{127}$I ratio of (2.0\,$\pm\,0.3)$\,x\,10$^{-13}$ was observed for the astro-grade, growth-grade and the ANU in-house analytical-grade powder, agreeing with the previously measured ratio by DAMA/LIBRA \cite{dama_libra}. All samples indicate no significant background contribution to the dark matter experiments. Therefore, from an $^{129}$I background perspective, there is no distinction in quality between different NaI powders assessed here. 
\\
\\
The $^{210}$Pb measurements presented within this work studied the performance of three different lead compounds Pb$_3$O$_4$, PbO$_2$ and PbO. The charge state yields of the 2+, 3+ and 4+ charge states were investigated for extracted PbO$^-$ and PbO$_2^-$ ions at different terminal voltages between 0.5\,MV and 0.9\,MV. A terminal voltage of 0.9\,MV and a charge state 3+ with a yield of 29.2\% from extracted PbO$_2^-$ ions were chosen for the data collection evaluated in this work due to the highest overall efficiency and minimising molecular ion interference. Furthermore, time evolution of the $^{208}$PbO$_2^-$ current from the investigated commercial powders was investigated, with PbO$_2$ and Pb$_3$O$_4$ both having high starting currents of about 450\,nA and PbO having a lower current of about 20\,nA. The overall measurement efficiency for $^{208}$Pb with charge state 3+ at VEGA at ANSTO for extracted PbO$_2^-$ ions was 1.3\,x\,10$^{-3}$.
\\
\\
Given that the expected $^{210}$Pb concentration in dark matter NaI(Tl) crystals is 0.03\,mBq/kg of NaI \cite{antonello}, the desired $^{210}$Pb/Pb isotopic ratio of the lead carrier when adding 1\,mg of stable lead into 1\,kg NaI(Tl) powder should be lower than 1\,x\,10$^{-14}$. The $^{210}$Pb/Pb isotopic ratio for the measured lead compound Pb$_3$O$_4$ was found in this region.  Our data indicate a machine background at VEGA for $^{210}$Pb measurements significantly lower than 10$^{-14}$ for $^{210}$Pb/Pb measurements. The investigated Pb$_3$O$_4$ could potentially be used as lead carrier. Therefore, within this work, we have demonstrated the capability to measure $^{210}$Pb at the required levels to characterise potential Pb carriers. However, an ideal lead carrier should have an $^{210}$Pb/Pb isotopic ratio  at least an order of magnitude lower than 1\,x\,10$^{-14}$. Lead materials from aged constructions were recently obtained, in which $^{210}$Pb should have significantly decayed, and hence could provide an ideal source of lead carrier material. These materials are currently being investigated.

\section*{Declaration of Competing Interest}
The authors declare that they have no known competing financial
interests or personal relationships that may influence
the work reported in this paper.

\section*{Acknowledgements}
The Heavy Ion Accelerator Facility (HIAF) at the ANU and the Centre for Accelerator Science at ANSTO are supported by the National Collaborative Research Infrastructure Strategy (NCRIS) of the Australian Government.


\end{document}